# Polarization and resistive switching in epitaxial 2 nm Hf$_{0.5}$Zr$_{0.5}$O$_2$ tunnel junctions

*Milena C. Sulzbach[†], Huan Tan[†], Saúl Estandía[†], Jaume Gàzquez[†], Florencio Sánchez[†], Ignasi Fina[†]\*, and Josep Fontcuberta[†]\**

[†] Institut de Ciència de Materials de Barcelona (ICMAB-CSIC), Campus UAB, Bellaterra, Catalonia 08193, Spain

*E-mail: ifina@icmab.es

*E-mail: fontcuberta@icmab.cat

KEYWORDS: ferroelectric tunnel junctions, hafnium oxides, Hf$_{0.5}$Zr$_{0.5}$O$_2$, resistive switching, tunnel electroresistance

ABSTRACT

In the quest for reliable and power-efficient memristive devices, ferroelectric tunnel junctions are being investigated as potential candidates. CMOS-compatible ferroelectric hafnium oxides are at the forefront. However, in epitaxial tunnel devices with thicknesses around ≈ 4 - 6 nm, the relatively high tunnel energy barrier produces a large resistance that challenges their implementation. Here, we show that ferroelectric and electroresistive switching can be observed in ultrathin 2 nm epitaxial Hf$_{0.5}$Zr$_{0.5}$O$_2$ (HZO) tunnel junctions in large area capacitors (≈ 300 µm$^2$). We observe that the resistance area product is reduced to about 160 Ω·cm$^2$ and 65 Ω·cm$^2$ for OFF and ON resistance states, respectively. These values are two orders of magnitude smaller than those obtained in equivalent 5 nm HZO tunnel devices while preserving a similar OFF/ON resistance ratio (210 %). The devices show memristive and spike-timing-dependent plasticity (STDP) behavior and good retention. Electroresistance and ferroelectric loops closely coincide, signaling ferroelectric switching as a driving mechanism for resistance change.

INTRODUCTION

Logic and memory elements in current computational architectures are spatially separated and settle a performance bottleneck. New devices and materials are intensively investigated to overcome this obstacle by integrating logic and storage functions in a single element.[1] Among them, memristors show characteristics that make them ideal for a new generation of computation[2] by emulating neuromorphic networks.[3] Memristors are two-terminal devices able to store multiple resistance states tuned by applying appropriate writing voltage pulses (electroresistance, ER). Different resistive switching mechanisms can give rise to ER. Anion or cation migration and phase change memories have been widely investigated.[4] However, these switching mechanisms present severe disadvantages, mostly related to their limited reliability and high power consumption.[5] In contrast, a resistive switching mechanism operating basically in open circuit mode, where only displacive electronic currents were at play, would lead to larger endurance and power benefits. Ferroelectric tunnel junctions fulfill these requirements and are receiving great attention.[5] Indeed, ferroelectric polarization switching is driven by external applied electric fields not requiring charge injection, and the subsequent modulation of energy barrier properties in tunnel devices leads to non-volatile changes in their conductance. Archetypical perovskite oxides (BaTiO$_3$, PZT, PbTiO$_3$, BiFeO$_3$, etc.) have been proved to show electroresistance (ER) driven by ferroelectric switching.[6-17]

The ferroelectric response discovered in CMOS-compatible Si-doped hafnium oxide films[18] and related compounds (Hf$_{1-x}$Zr$_x$O$_2$ and others) has attracted huge interest.[19-20] Contrary to conventional ferroelectrics, the ferroelectricity in hafnium oxide films is enhanced at the ultrathin limit (thickness < 10 nm). It has been proposed that surface energy boundary conditions largely affect the subtle balance between the energy of competing polymorphs (monoclinic, orthorhombic, tetragonal, or cubic phase) of HfO$_2$, favoring the ferroelectric orthorhombic phase only in a restricted thickness range.[21] In ferroelectric ultrathin films, a rhombohedral distortion has also been observed.[22] The discovery of ferroelectricity in $HfO_2$ nanometric films has stimulated the research to understand and control any connection between polarization switching and related ER phenomena.[23-34] However, as early pointed out by Max et al.,[35] and in agreement with experimental observations in other ferroelectrics,[36-38] the coexistence of ferroelectricity and ER in a $HfO_2$-based device does not guarantee their causal connection. Indeed, non-ferroelectric $HfO_2$ films display ER due to filamentary conductivity and ionic motion-related mechanisms,[39] leading to resistive switching[5].

In ultrathin hafnium oxide-based films, the reported resistance ratio between OFF/ON states (= $R_{OFF}/R_{ON}$),[23, 26-28, 35, 40-44] range from 300% [42] to 5000%,[40-41] suggesting that different mechanisms may contribute to ER. Therefore, disentangling these seemingly similar responses of disparate origins is challenging, particularly in polycrystalline hafnia films. Indeed, even the physical origin of the ferroelectric behavior of the HZO films is still under debate. Recent results suggest that electrochemical processes coupled to strain can trigger the observation of a switchable polarization.[45-46]

Epitaxial growth of hafnium oxide films[22, 47, 48] has been demonstrated to be instrumental in identifying and control displacive and ionic contributions to ER.[40-41, 49-50] The improved crystalline quality of epitaxial films helped to mitigate the ionic conductivity and associated ER, widely studied in polycrystalline $HfO_2$[51-54] and $ZrO_2$[55] films, which are known to be governed by existing grain boundaries. Indeed, epitaxial engineering[56] allows to control the presence of grain boundaries in $Hf_{0.5}Zr_{0.5}O_2$ (HZO) films and understand their contribution to ferroelectric properties and ER, leading to its near-complete suppression by appropriate substrate selection.[56] However, epitaxial hafnium oxide ferroelectric tunnel junctions, typically (4 - 8) nm thick, show extremely high resistance values. The reason is intrinsic: hafnium oxide ($Hf^{4+}$:$5d^06s^0$) and zirconium oxide ($Zr^{4+}$:$4d^05s^0$) have small electron affinity.[57-60] It follows that the tunnel barrier height in a metal-ferroelectric-metal junction will be larger in a hafnium oxide than in oxide perovskite ferroelectrics.[61] This results in low tunneling current and high resistance values. Whereas high resistance is desirable for some applications, for instance, in ferroelectric complementary resistive switching devices,[62-63] a sizable current is required to achieve fast and reliable reading operation of memory devices, and a trade-off is to be found.[5]

Different strategies can be followed to reduce device resistance: first is to use metals with lower work function to reduce tunneling barrier height (at expenses of usually higher reactivity of the metal); second is to stimulate Fowler-Nordheim tunneling (at expenses of requiring higher reading voltages);[64] and third is to reduce the barrier thickness. The recent discovery of ferroelectricity and ER in 1 nm HZO films[43] and the observation of robust ferroelectric properties in sub-5 nm HZO capacitors[65-66] suggest that extreme ultrathin HZO may offer the optimal solution.

Here, ferroelectric HZO films down to 2 nm are grown by pulsed laser deposition (PLD), avoiding the fast thermal annealing post-deposition and metal capping steps required to promote the formation of the ferroelectric o-HZO in atomic layer deposition (ALD) grown films.[20] Moreover, we use scandate single-crystalline substrates that boost the selective growth of the ferroelectric o-HZO polymorph, thus minimizing grain boundaries in the film.[56] We show that ferroelectricity coexists with tunnel ER in HZO films as thin as 2 nm, in large capacitor structures ($\approx$ 20 µm diameter) and we prove their univocal relationship. Polarization reversal is directly probed, and it is accompanied by a large ER that reaches about 210 % at room temperature. As expected, the resistance is found to be reduced by about two orders of magnitude compared with 5 nm tunneling barriers. We demonstrate memristive, spike-timing-dependent plasticity (STDP) behavior and good retention, with switching time below 500 ns in these extreme ultrathin films.

RESULTS AND DISCUSSION

Epitaxial HZO films with 2, 3, and 5 nm thickness were grown on $GdScO_3$(001)-oriented (GSO) single crystalline substrates ($GdScO_3$ is indexed in pseudocubic setting), using $La_{2/3}Sr_{1/3}MnO_3$ (LSMO) (22 nm thick) as bottom conducting electrode, by pulsed laser deposition (PLD) as described elsewhere.[56]

In **Figure 1**a, we show voltage-dependent piezoresponse force microscopy (PFM) amplitude and phase for a pristine sample of the 2 nm HZO collected by placing the tip on the bare surface of HZO. Phase loops indicate a phase change of 180° upon poling, and the PFM amplitude is almost zero at the coercive voltage, as expected for an intrinsic ferroelectric response. Data collected on Pt electrodes show similar results (Figure S1). PFM loops collected after repeatedly applying voltages up to ± 8 V ($N_C$ = 100 cycles), also included in Figure 1a, show similar behavior. The only perceptible differences are i) a minor increase of the piezoelectric amplitude response and coercive voltage and ii) a decrease of the imprint. Data collected for HZO (3 and 5 nm) films display the same features (Figure S2). Figure 1b shows PFM phase maps of the surface of a pristine HZO (2 nm) sample after electrical poling using - 6 V (dark regions) and + 6 V (bright regions). Figure 1c shows a PFM image collected after-field cycling ($N_C$ = 3) in the explored region before final ± 6 V voltage poling. The close similarity of images and PFM loops in pristine and field-cycled samples strongly indicate that field cycling does not significantly modify the ferroelectric character of the films. Outer bright regions have not been poled, and images indicate that the as-grown polarization state is downwards in both samples. Inset shows a different pattern also obtained for ± 6 V voltage poling. Amplitude and phase images shown in Figure S3 (2 and 3 nm films) show 180º phase shift and near-constant amplitude (zeroing at the domain walls) between regions poled with opposite voltage, suggesting that charging contribution are not prevalent.[67]

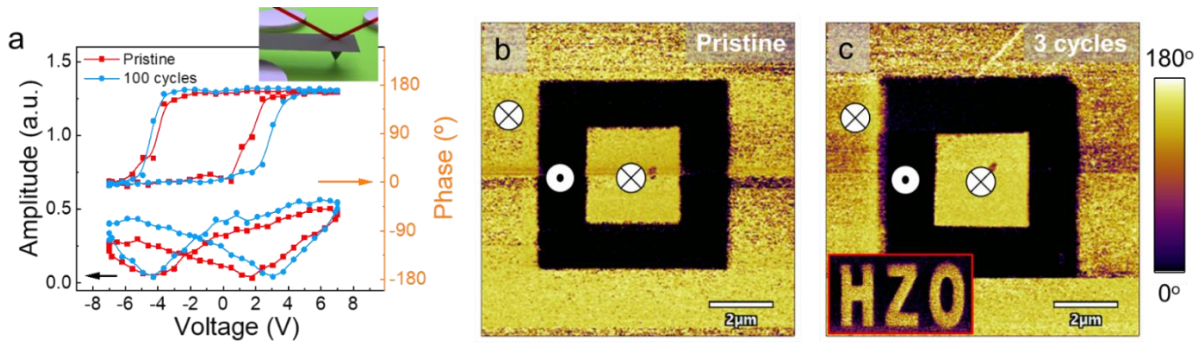

**Figure 1.** a) Phase and amplitude PFM loops of HZO (2 nm) pristine and voltage-cycled ($N_C$ = 100) sample. Inset: illustration of signal (laser) collection and monitoring cantilever deformation. PFM phase images collected in the region (10 x 10) µm² for b) pristine sample and c) after cycling ($N_C$ = 3). Bright and dark regions were obtained by poling with + 6 V and - 6 V, respectively. The inset in c) shows (5 x 10) µm² PFM phase map collected in a region poled with a different ("HZO") pattern.

**Figure 2**a (background image) shows X-ray diffraction 2θ-χ frame of the HZO (2 nm) film. The diffraction spots corresponding to (001) and (002) reflections of substrate and LSMO and those of orthorhombic (o-) or rhombohedral distortion HZO are well visible. Note that available data does not allow to distinguish between these two structures.[56] In Figure 2a, we also include the integrated scan along χ (+/- 10°); In the 20° < 2θ < 50°

angular range, which is dominated by the strong diffraction spots of the substrate (GSO) and bottom electrode (LSMO). Zoom in the 20° < 2θ < 42° range allows identifying the diffraction spots of o-HZO. The diffraction peak of HZO narrows, and its intensity increases for increasing HZO thickness (Figure S4 and Figure S5). Data display the obvious presence of o-HZO (111) reflection at 2θ ≈ 30° and the absence of the m-HZO (002). Further insight into the HZO (2 nm) microstructure was obtained by scanning transmission electron microscopy (STEM). The contrast of the high angle annular dark-field (HAADF) imaging mode, scaling approximately as the square of the atomic number Z of each atomic column, permits the observation of the cations sublattices in the HZO as in the LSMO films. Figure 2b shows a high-magnification STEM-HAADF image of a pristine HZO (2 nm) sample, where an isolated m-HZO grain is observed next to several o-HZO grains. The image in Figure 2b demonstrates the presence of a minor fraction of m-HZO but does to provide a quantitative estimation of its relative fraction in the film. Indeed, m-HZO was observed only in some frames. The region marked with a yellow square, shown enlarged in Figure 2c, identifies a (002) textured m-HZO grain next to a (111) oriented o-HZO one. Therefore, HAADF images reveal that a large majority of o-HZO grains seldom coexist with m-HZO grains. These results, together with the fact that XRD 2θ-χ frames and 2θ scan (Figure 2a) do not show a distinct monoclinic reflection, indicate that the amount of m-HZO phase is rather small.

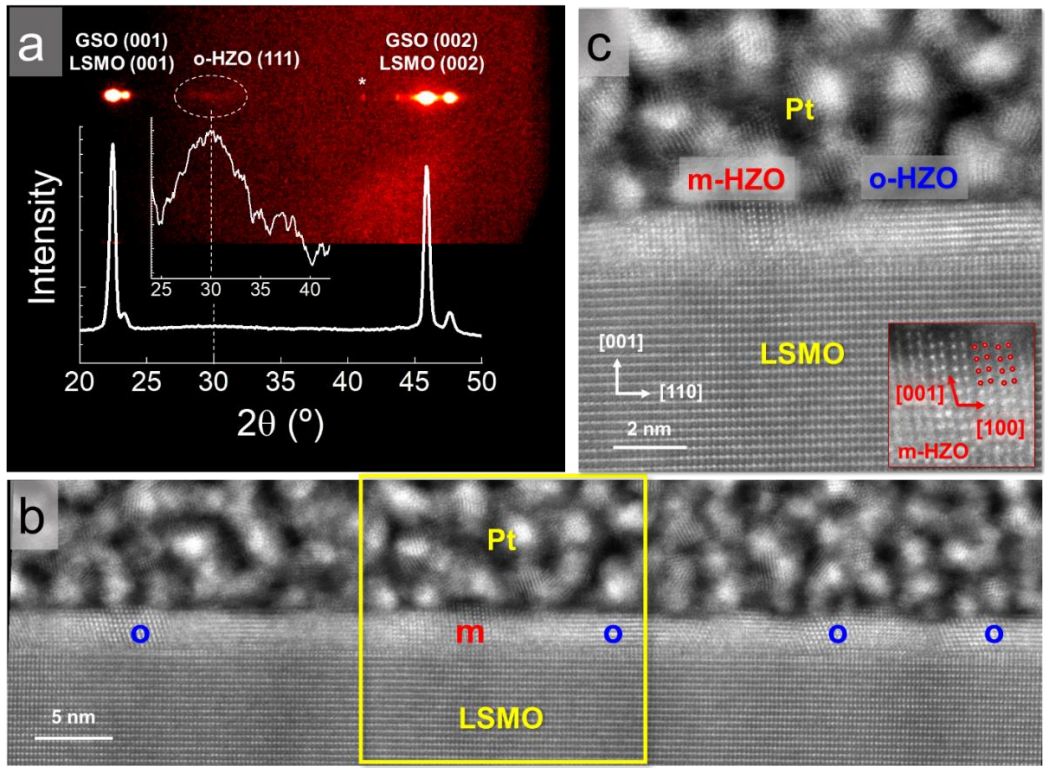

**Figure 2.** a) XRD 2θ-χ frame and integrated scan of the sample with the 2 nm HZO film. Integrated scan along χ and corresponding zoom around o-HZO (111) is also included. The tiny bright spot marked with an * is an artifact from the detector. b) Cross-sectional HAADF-STEM image of the HZO/LSMO heterostructure, where one m-HZO grain is indicated next to two o-HZO. c) Enlarged view of marked region in which m-HZO (red "m") grain and o-HZO (blue "o") grains are indicated. The inset in c) shows the monoclinic grain with the simulated cation sublattice of the monoclinic phase (space group $P2_1/c$) superimposed.

Circular Pt top electrodes of 20 µm of diameter were grown at room temperature through shadow masks by sputtering. The sample is sketched in **Figure 3**a. Figure 3b shows illustrative I(V) curves and the corresponding polarization P(V) loops of the 2 nm HZO film recorded at 5 kHz after $N_C = 10^4$ training cycles using Positive-Up-Negative-Down (PUND) technique (see Methods). The current peaks associated with polarization switching are evident, and the coercive voltage extracted from the switching peak positions is $V_c \approx 2.8$ V; the presence of a small imprint $V_{imp} = + 0.1$ V can also be observed. Note that here the imprint is different from PFM data ($V_{imp} = - 0.6$ V, Figure 1a), probably due to the large number of cycles performed before the P(V) loop measurement (Figure 3b). The integration of the current indicates a remnant polarization of $P_R \approx 2$ µC·cm$^{-2}$. PUND data before and after subtraction clearly shows a large leakage current superimposed to the displacive current occurring at the coercive field (Figure S6). Before cycling, the displacive current is hidden by the larger leakage current (Figure S7).

Figure 3c shows the reading I(V) curves corresponding to two resistance-state after pre-polarizing the junction with $V_w = \pm 5$ V and $\tau_w = 300$ µs. It can be seen that the curves are sigmoidal and slightly asymmetric, as commonly found in trapezoidal tunnel barriers. Importantly, for $V_w = -5$ V (down triangles), the conductance is larger than for $V_w = +5$ V (up triangles) (see also Figure 1c (inset)), indicating that the writing voltage $V_w$ sets the junction resistance. Figure 3d shows the electroresistance (ER) $R(V_w)$ loops collected with $-V_{max} < V_w < V_{max}$, with $V_{max} = 5$ V (green symbols) after $N_C = 10^4$ cycles. It is observed that the resistance rapidly changes at around $\pm 3$ V, leading to well-defined low/high (ON/OFF) resistance states of the junction. If $V_{max}$ is increased to 8 V (purple symbols), similar behavior is observed, also leading to similar although distinct ON/OFF states. Five $R(V_w)$ consecutive loops for every $V_{max}$ are shown to illustrate the robustness of the measured ER. Similar loops are obtained using lower reading voltage (Figure S8). The extracted OFF/ON ratio is 150 % for $V_{max} = 5$ V and 210 % for $V_{max} = 8$ V. These values are comparable to those reported for ER associated with genuine polarization switching in similar tunnel barriers (410% [49] and 400% [40-41]). Larger ER had been reported in some HZO barriers, either not related to tunnel transport (16100% [68]) or attributed to ionic motion or after repeated voltage cycling of the junctions ($10^5$ % [49] - $10^6$ % [40-41]). The close coincidence between the coercive voltages where polarization switches (Figure 3b) and the $V_w$, where the resistance starts to change in Figure 3d (yellow line), suggests that the ER is here due to polarization-controlled tunnel transport across the 2 nm HZO barrier. Consistently, the derivatives d(P(V))/dV and d(ER(V))/dV display closely coincident maxima at similar values $(V = V_C)$ (Figure S9), as expected from a common origin.[35-36] Note that the presence of leakage current across the tunneling barrier might a priori hamper polarization switching and the subsequent ER. Therefore, the observation of the resistive switching coinciding with the polarization reversal is remarkable by itself. The observation of a prevalent polarization-related ER rather than the ionic one can be ascribed to the lower abundance of grain boundaries and associated defects in these epitaxial films, the use of a low reactive top electrode (here Pt) and/or the reduced thickness compared with other ferroelectric hafnium oxide barriers.[23, 26-

[28, 35, 40-44] Following Cheema et al.,[43], we also used voltage-polarity-dependent current-voltage hysteresis to disregard resistive switching mediated by the dielectric breakdown and filamentary-type switching as a source of the observed ER (see Figure S10). As shown in Figure 3d, the ER loops of pristine samples indicate a lower resistance ($\approx 3 \cdot 10^5$ Ω), which is about two orders of magnitude smaller than after $N_C = 10^4$ cycles, and have a reduced ER ($\approx 1$ %). Detailed analysis of the I(V) characteristics allows concluding that voltage training favors the blocking of non-tunnel transport channels across the HZO barrier (Figure S11 and Table S1). Based on the available information, it cannot be excluded that the voltage cycling effect might be related to the common "wake-up" in ferroelectric $HfO_2$ involving a field-induced phase transformation of the residual m-HZO to o-HZO. As reported elsewhere,[49] coexistence of m-HZO and o-HZO phase leads to HZO barriers of smaller resistance.

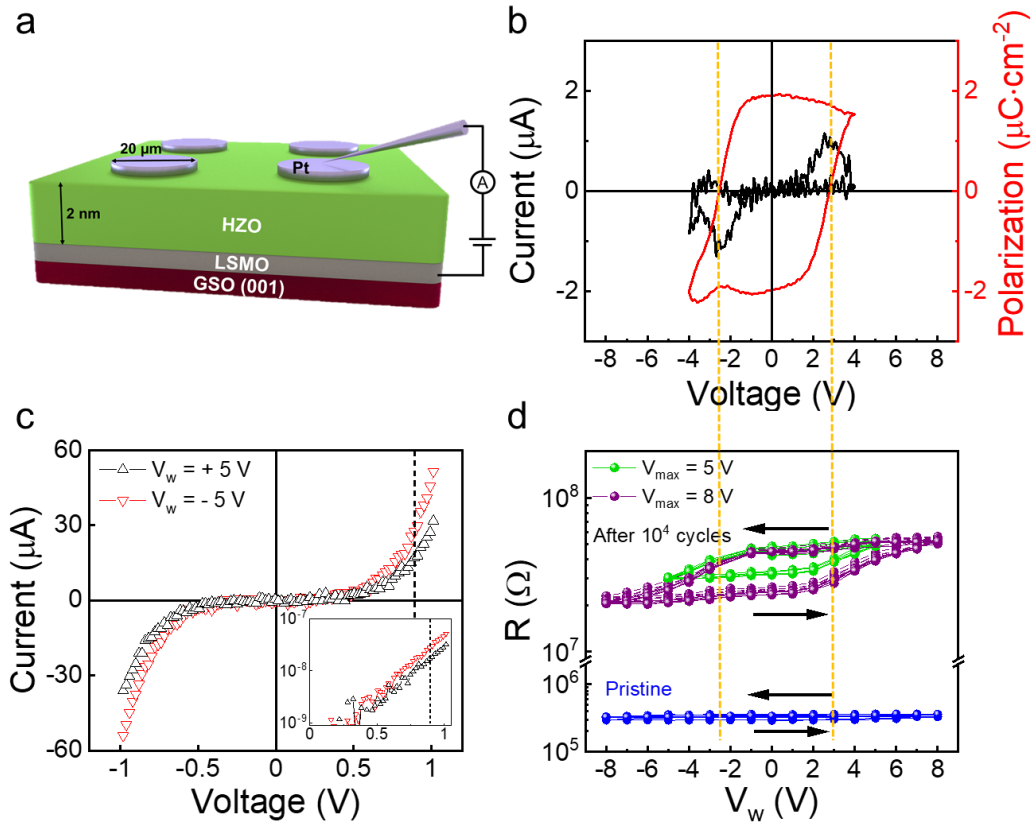

**Figure 3.** a) Sketch of the sample. b) Current-voltage I(V) and polarization loop ($V_{max}$ = 4 V) from 2 nm junction measured at 5 kHz after $10^4$ training cycles. c) I(V) curves collected after writing with $V_w$ = + 5 V and $V_w$ = - 5 V as indicated. Inset: log scale of current (units as in the main panel). d) Dependence of junction resistance on the writing voltage measured in pristine junction and after $10^4$ cycles. The junction is the same as in (b). Five sequential loops are shown for each $V_{max}$. Dashed vertical lines in b) and c) indicate the coercive voltage ($V_c$).

The ER data in Figure 3d is characteristic of memristors and indicates that the device resistance could be tunable. We emphasize that the tunnel devices reported here are HZO (2 nm) insulating layers in large area capacitor structures ($\approx$ 300 µm$^2$). In order to explore the memristive behavior in more detail, the resistance state has been tracked while applying a series of $N_p$ writing pulses with the same amplitude and polarity, mimicking potentiation/depression processes in neuronal firing. **Figure 4**a shows the dependence of the resistance on the number of writing pulses ($R(N_p)$) of each polarity, as sketched in the bottom panel. It can be appreciated that the resistance of the device changes gradually with $N_p$. The percentage of resistance increase or decrease depends on the amplitude of the $V_w$ pulse and its duration. In Figure 4b we used: $V_w$ = ± 5V and ± 8 V, and $\tau_w$ = 3 µs and 300 µs. Indeed, both longer time and larger amplitude pulses enhance the change of resistance. Time-dependence switching characterization shows that these observations are limited by the time constant of the device under test (the ferroelectric capacitor and the series resistance of the bottom electrode) and measuring circuit (≈ 500 ns) (Figure S12). The reproducibility of this behavior for eight cycles is demonstrated in Figure 4b, which also proves the robustness of the memristive response. Consistent with earlier findings,[69-70] the potentiation/depression data fit well into an exponential model (Figure S13 and Table S2). The endurance of the junction electroresistance (Figure S14) is similar to that reported in hafnia films (polycrystalline) of similar thickness.[71] In addition, these ultrathin junctions have remarkable retention time (non-volatile data storage). Figure 4c shows the time evolution of the ON/OFF resistance states, set by applying writing pulses of $V_w = \pm 5$ and $\pm 8$ V and $\tau_w = 300$ μs, that persist for at least $10^3$ s. It can be appreciated in Figure 4c that the resistances for $V_w = + 8$ V and +5 V are similar, but in the case of $V_w = - 8$ V, the resistance is smaller than for $V_w = - 5$ V. A related asymmetric ON/OFF switching have been observed in other ferroelectric based tunneling junctions.[6, 72-74] Wen et al.[73] attributed the asymmetric ON-to-OFF and OFF-to-ON behavior to the formation of space charge regions at the oxide-metallic electrode interface. Chanthbouala et al.,[72] in contrast, attributed the asymmetry to the distinct nucleation of up/down domains and subsequent propagation ruled by the Kolmogorov–Avrami–Ishibashi (KAI) model, which is the usual case for clean epitaxial systems.[75] A similar scenario may also hold here. We note that in polycrystalline $HfO_2$-based ferroelectrics, where a distribution of local switching times ($t_0$) exists, the switching process is better described by the Nucleation Limited Switching (NLS) or other models.[76-77] The epitaxial nature of our films seems to narrow the $\{t_0\}$ distribution, thus leading to the limiting (KAI) model, corresponding to a polarization reversal via homogeneous polar regions nucleation and domain-wall motion. More importantly, the ON/OFF states do not show any degradation within the explored time window ($10^3$ s).

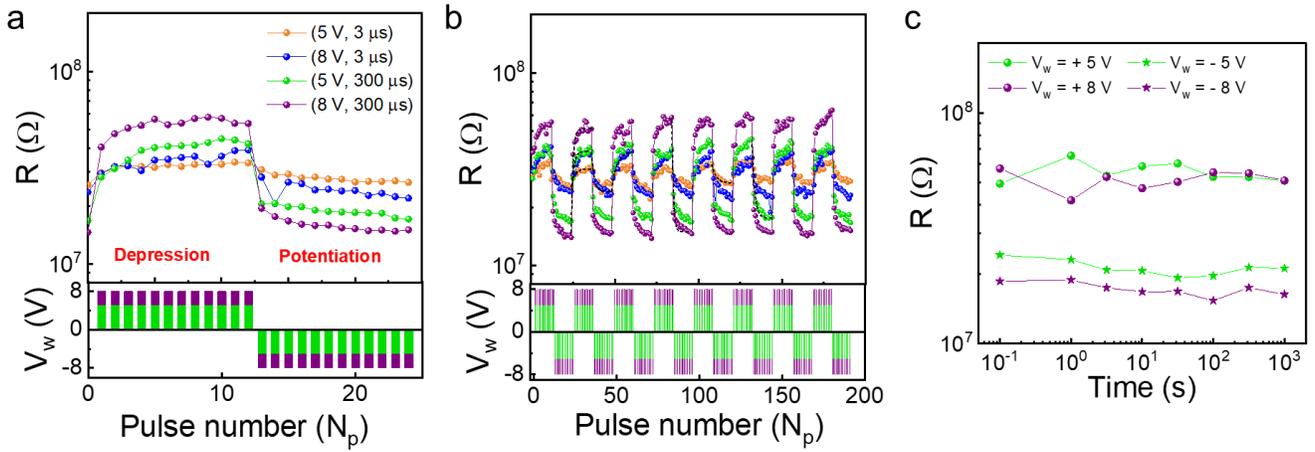

**Figure 4.** a) Memristive behavior measured in 2 nm HZO FTJ (top panel) by applying $N_p = 12$ pulses of different polarity (bottom panel) and duration as indicated. b) Resistance for the repetition of measurement in (a) for eight potentiation/depression cycles (top panel) and the corresponding voltage pulses (bottom panel). c) Retention after writing pulse with indicated amplitude and 300 µs duration.

In recent years, efforts have been made to design and integrate devices that mimic synaptic characteristics into brain-inspired systems. Biological synapses connect different neurons and can modulate their conductivity through their plasticity, which is associated with their learning and adaptation abilities. Artificial synapses can be modeled by memristors, i.e., resistors that change their resistance state depending on the voltage inputs from the so-called pre- and post-neurons.[78] Spike-timing-dependent plasticity (STDP) is the model that correlates the modulation of the synapse conductance (state) with the time difference between the firing of the pre- and post-neurons. (**Figure 5**a). We have performed an STDP characterization in the HZO (2 nm) tunnel junctions to evaluate the device plasticity under similar conditions to which biological synapses are exposed. When resistive switching devices are used to explore synaptic-like plasticity, two different electrodes are commonly used to apply $V_{pre}(t)$ and $V_{post}(t+\Delta t)$ voltages pulses separated by a time shift $\Delta t$, mimicking the pre and post-synaptic action potentials. Then, the junction's resistance or conductance is measured as a function of $\Delta t$. Here, for the sake of simplicity, time-shifted spikes of opposite polarity are generated and added by a suitable power supply (Figure 5c) and applied to the top Pt electrode, while the LSMO bottom electrode remains grounded (Figure 5b). The conductance ($\Delta G(\Delta t) = (G(\Delta t)-G(\Delta t=100\ \mu s))/G(\Delta t=100\ \mu s)$) for a 2 nm junction after each voltage spike dependence is displayed in Figure 5d. The modulation of conductance found in our devices has a biologically realistic STDP shape.[78] It can be observed that the conductance can be modulated by ≈ 60 %. The observed G(Δt) modulation ratio is similar[68] to or even larger[70] than earlier reported on HZO[74, 79] ferroelectric tunnel barriers that show much larger resistance (≈ $10^5$ Ω·$cm^2$) than the barriers reported here. Fittings of ΔG(Δt) to equation ΔG(Δt) = A·exp((Δt)/ (Δ$t_0$)) (solid lines in Figure 5d) indicate characteristic times Δ$t_0$ = 64 and 14 μs for positive and negative time-shifts. These values are in agreement with values reported in ultrathin $Hf_{0.5}Zr_{0.5}O_2$ based junctions.[79-80] The asymmetry of the characteristic times Δ$t_0$, commonly observed in ferroelectric devices, reflects the asymmetric switching of the ferroelectric polarization, as commonly observed.

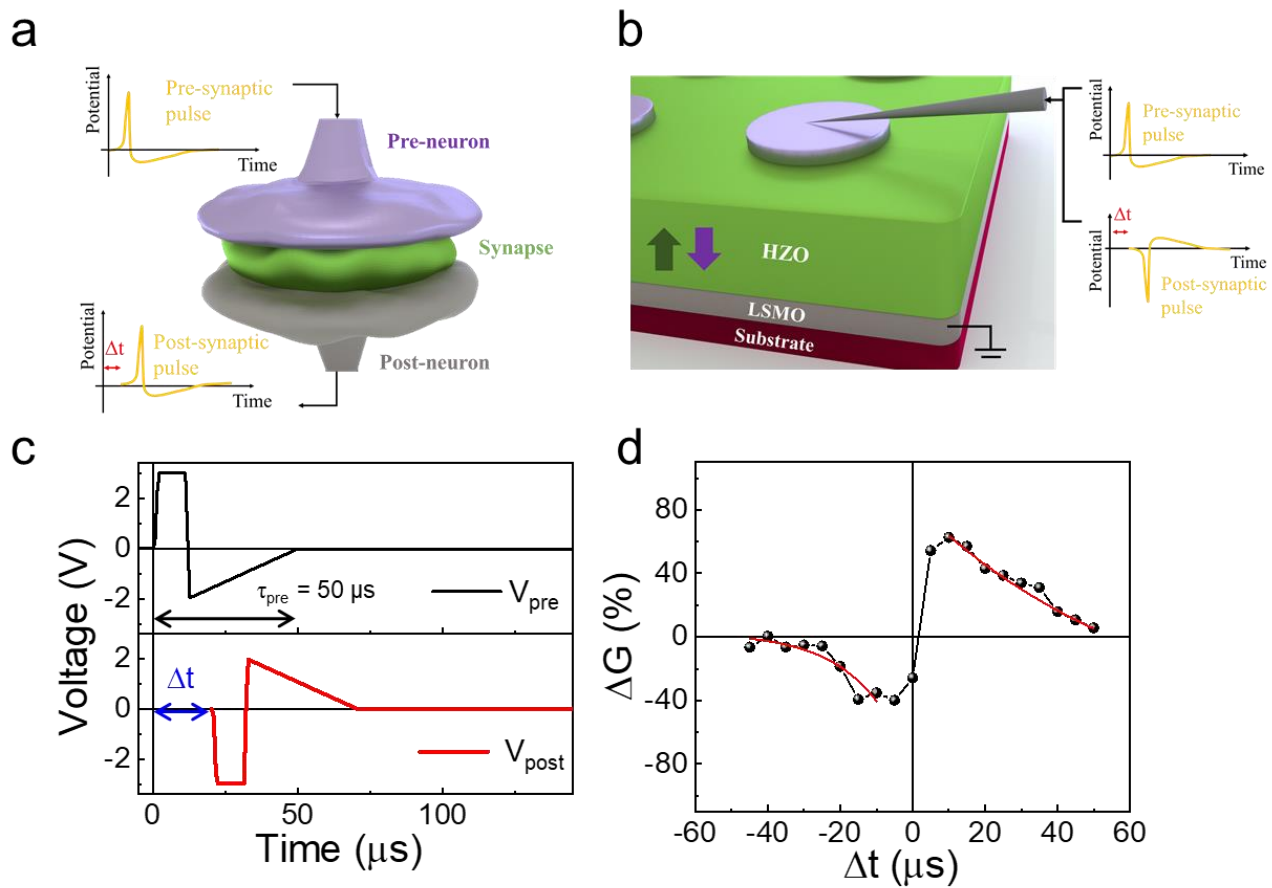

**Figure 5.** a) Sketch of the synapse and the pre-neuron and post-neuron connection. b) Sketch of the implementation of the spike-timing-dependent plasticity (STDP) characterization in the characterized devices. c) Sequence of waveforms (spikes) applied on top electrode while bottom electrode remains grounded for STDP measurement. Δt = $t_{pre}$ - $t_{post}$, where $t_{pre}$ and $t_{post}$ are the times corresponding to the signal trigger. Both spikes have a duration of 50 μs, as indicated in the top panel. d) STDP measurement showing modulation of conductance (ΔG) as a function of Δt. Lines across the data points correspond to the fitting of the exponential function ΔG(Δt) = A·exp((Δt)/(Δ$t_0$)) within the [10 and +50 μs] and [-10 and -50 μs] time intervals.

Having analyzed the structure, polarization, and ER of the ultrathin HZO (2 nm) films, we performed a similar analysis on thicker HZO films (3 and 5 nm). PUND data (Figure S15) shows $P_r \approx 5$ µC·cm$^{-2}$ and $P_r \approx 25$ µC·cm$^{-2}$ for the 3 and 5 nm HZO films, respectively, illustrating that $P_r$ displays a peak at a certain thickness, as commonly found in epitaxial[65, 81-82] and polycrystalline films.[83-87] It has also been observed that the piezoelectric amplitude also increases with thickness up to 5 nm (Figure S16). Note that a piezoelectric-like response can result from the large electrostriction inherent in fluorite lattices with sizeable content of oxygen vacancies[88] and their motion under electric field. However, available models[45] predict a monotonic thickness dependence of the polarization, which is not observed in the present case. The corresponding ER loops displayed in **Figure 6**a for HZO (3 nm) show an increased value on the OFF/ON ratio of 145 % ($V_{max}$ = 5 V) and 280 % ($V_{max}$ = 8 V), respectively. It is worth noticing that the number of cycles required to obtain a sizable and stable ER reduces when increasing thickness. In the case of HZO (5 nm) films, the resistance loops (Figure 6b) display an OFF/ON ratio of 130 % for $V_{max}$ = 5 V and 200 % for $V_{max}$ = 8 V only after 8 cycles, which are similar to those recorded in the pristine state (Figure S17) and fully consistent with reported ER data in HZO films of nominally identical thickness.[49-50] In short, these observations indicate that voltage cycling is only relevant to reduce current leakage and obtain stable ER in HZO films thinner than 3 nm, probably reflecting that interfacial carrier accumulation and defects are seemingly more relevant in the ultrathin films. Figure 6c summarizes the resistance of junctions of different thicknesses in their pristine state and after voltage cycling as a function of their thickness. For pristine 2 and 3 nm devices, the resistance is R $\approx 3 \cdot 10^5$ and $\approx 4 \cdot 10^7$ Ω, respectively, in the pristine state and, as mentioned, there is no significant ER. After cycling, the junction resistance increases, and ER emerges. Importantly, Figure 6c emphasizes that cycled HZO (2 nm) junctions have a resistance about 2 orders of magnitude smaller than HZO (5 nm) with a similar ER response. This large ER is obtained in ultrathin tunnel devices with a resistance area product $\approx 160$ Ω·cm$^2$ and 65 Ω·cm$^2$, which is several orders of magnitude smaller than previously found in polycrystalline HZO ferroelectric tunnel junctions measured at a similar voltage ($\approx 10^4 - 10^6$ Ω·cm$^2$).[23, 30, 42, 74, 79, 89] In epitaxial films, similar resistance area product values have been reported, although conductance and ER of the junctions have been partially assigned to ionic motion.[40-41] Interestingly, in 1 nm HZO films (75 % content of Hf)[43] directly growth on Si or in TiN/HZO(2.5 nm)/Pt[26] structures, where ER has been attributed to ferroelectric switching, similarly small resistance values have been reported.

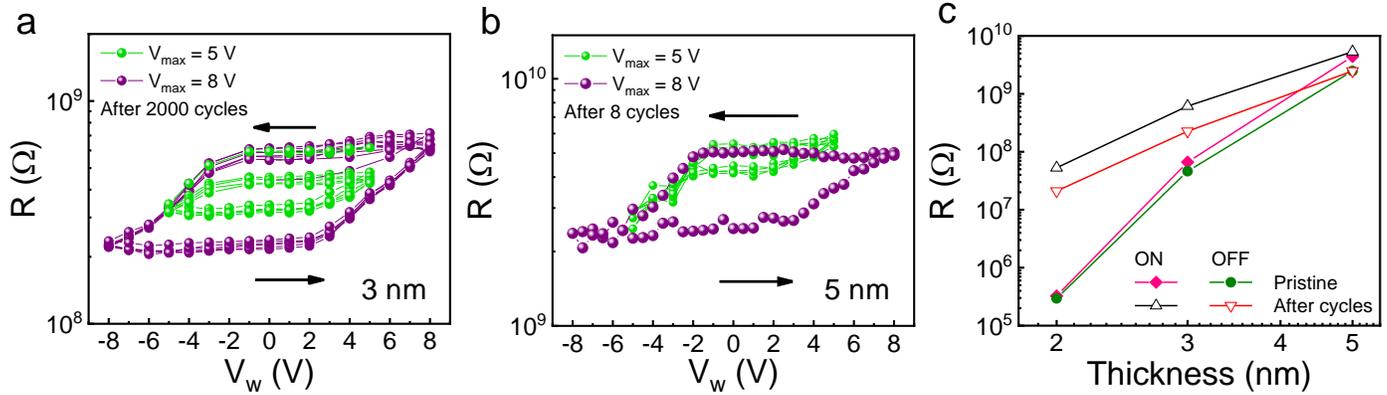

**Figure 6.** Resistance dependence with writing voltage obtained for a) 3 nm and b) 5 nm FTJ after cycling. In both cases, the pulse width was 300 µs. c) ON and OFF resistance states dependence on the HZO thickness for pristine and cycled junctions.

CONCLUSIONS

We have reported here the observation of ferroelectric and piezoelectric polarization loops and ER in large capacitor structures ($\approx 300$ µm$^2$) containing ultrathin (2 nm) epitaxial layers of HZO. The ultrathin HZO films are grown on GdScO$_3$ and are mainly constituted by the ferroelectric o-HZO phase and display well-behaved piezoelectric and ferroelectric responses in PFM. The close coincidence between coercive fields extracted from P(V) and ER loops and the PFM piezoelectric loops indicate the connection between ferroelectric polarization switching and the electroresistance, which reaches up to 210 %. Interestingly, this large ER is obtained in ultrathin tunnel devices with a reduced resistance area product $\approx 160$ $\Omega \cdot$cm$^2$ and 65 $\Omega \cdot$cm$^2$ for OFF and ON resistance states. Multi-resistive states can be obtained *on demand* by controlling the number of applied voltage pulses and their duration, and this feature is used to demonstrate conductance modulation as large as 60 % in STDP measurements. The switching time (limited by the sample and the measuring experimental set-up time constant) is below 500 ns. In the pristine state, these ferroelectric large area capacitors display a substantial leakage current that precludes observation of polarization switching and electroresistance. Upon cycling, leakage is suppressed, the resistance increases, and polarization switching and ER emerge. It is found that when increasing the HZO thickness, the voltage cycling is gradually less relevant and, for 5 nm, no longer required. Observation of long data retention in ultrathin HZO tunnel barriers with reduced resistance area product shall contribute to their exploitation in functional devices.

EXPERIMENTAL

*Samples growth.* Epitaxial HZO films with 2, 3, and 5 nm thickness were grown on $GdScO_3$ (001)-oriented (using pseudo cubic setting) (GSO) single crystalline substrates (5 x 5 mm$^2$) buffered with $La_{2/3}Sr_{1/3}MnO_3$ (22 nm thick) conducting electrodes by PLD, as described elsewhere.[56] Circular Pt top electrodes of 20 µm of diameter and 20 nm in thickness were grown through shadow masks by sputtering.

*Electrical characterization.* Electrical characterization was performed by contacting the bottom electrode and one top Pt electrode. All electric characterization was performed using the TFAnalyser2000 platform (Aixacct Systems Gmbh.). The resistance is measured after each writing pulses ($V_w$) by applying a linear $V_R(t)$ triangular pulse in a small voltage range (from – 1 V to + 1 V) after a delay time of $\tau_D = 0.2$ s. Note that the maximum reading voltage is smaller than $V_c$ to avoid ferroelectric switching during the reading. The resistance is calculated at $V_R = 0.9$ V. The $V_w$ pulse is applied before resistance measurement using a trapezoidal signal with duration $\tau_w$. Further details are described elsewhere.[56] The junctions cycling (named training in the present work) was done using bipolar square pulses of indicated number at 1 kHz and 3 V amplitude. STDP measurements were done by digitally generating two signals ($V_{pre}(t)$ and $V_{post}(t)$) and then digitally adding with a varying time-shift ($\Delta t$), and used as an input of a signal generator (TFAnalyser2000). Each signal ($V_{pre}(t)$ and $V_{post}(t)$) was composed by a first positive square pulse of 2.5 V and 10 µs and an increasing ramp from -2 V to 0 V of 40 µs duration, as shown in Figure 5c. The output signal ($V_{Tot}(t) = V_{pre}(t) + V_{post}(t + \Delta t)$) is subsequently applied to the top electrode. Note that the polarity of $V_{pre}$ and $V_{post}$ is inverted to mimic the two signals of the same polarity separated by $\Delta t$ that are applied to the top and bottom electrodes in most common experimental arrangements.

*Structural characterization.* X-ray diffraction experiments were performed using a point detector in θ-2θ configuration using a Bruker-AXS (model A25 D8 Discover). 2θ-χ frames were recorded by a Bruker D8-Advance diffractometer equipped with a 2D detector.

*Scanning Transmission Electron Microscopy.* Microstructural characterization of the 2 nm thick HZO sample was performed by scanning transmission electron microscopy using a JEOL ARM 200CF STEM operated at 200 kV and equipped with a cold field emission source and a CEOS aberration corrector. The STEM specimen was prepared by a FEI Helios nanolab 650 focus ion beam. High-angle annular dark-field images of the cross-sectional specimen were recorded along the substrate's pseudocubic [110] zone axes.

*Piezoelectric Force Microscopy.* Piezoresponse force microscopy (PFM) measurements were performed using an MFP-3D ASYLUM RESEARCH microscope (Oxford Instrument Co.) with conductive Pt on both sides coated tips (Multi75E-G, BudgetSensors). In order to achieve better sensitivity, the dual AC resonance tracking (DART) method was employed.[90] PFM voltage hysteresis loops were always performed at remanence, using a dwell time of 5 ms.

ASSOCIATED CONTENT

**Supporting Information**. The following files are available free of charge.

Additional figures include PFM data, XRD 2θ-χ scans, I(V) and P(V) data, dP/dV and $dR/dV_w(V)$ loop, $R(V_w)$ loops, current-voltage hysteresis, description of I(V) fitting model, measurement of circuit time constant, fitting of potentiation/depression data, endurance measurements, resonance spectra collected with PFM (PDF).

AUTHOR INFORMATION

**Corresponding Author**

*E-mail: ifina@icmab.es (I.F.), fontcuberta@icmab.cat (J.F.)

**Notes**

The authors declare no competing financial interest.

ACKNOWLEDGMENTS

Financial support from the Spanish Ministry of Science and Innovation, through the Severo Ochoa FUNFUTURE (CEX2019-000917-S), MAT2017-85232-R (AEI/FEDER, EU) and PID2019-107727RB-I00 projects, and from Generalitat de Catalunya (2017 SGR 1377) is acknowledged. M.C.S. acknowledges the fellowship from "la Caixa Foundation" (ID 100010434; LCF/BQ/IN17/11620051). I.F. acknowledges Ramón y Cajal contract RYC-2017-22531. Project supported by a 2020 Leonardo Grant for Researchers and Cultural Creators, BBVA Foundation. M.C.S and S.E. work has been done as a part of her Ph.D. program in Physics at Universitat Autònoma de Barcelona. H.T. work has been done as a part of her Ph.D. program in Material Science at Universitat Autònoma de Barcelona. S.E. acknowledges the Spanish Ministry of Economy, Competitiveness and Universities for his PhD contract (SEV-2015-0496-16-3) and its co-funding by the ESF. HT is financially supported by China Scholarship Council (CSC) with No. 201906050014. Anna Crespi and Francesc Xavier Campos are acknowledged for valuable support on structural characterization. Electron microscopy observations were carried out at the ICTS-CNME at UCM, Madrid. The authors acknowledge the ICTS-CNME for offering access to their instruments and expertise. The STEM specimens were prepared in the Service of Dual Microscopy FESEM & FIB, being part of the Central Services for Research of the University of Málaga and placed in the Supercomputing and Bioinnovation Center.

# Supporting Information

Polarization and resistive switching in epitaxial 2 nm $Hf_{0.5}Zr_{0.5}O_2$ tunnel junctions

*Milena C. Sulzbach[†], Huan Tan[†], Saúl Estandía[†], Jaume Gàzquez[†], Florencio Sánchez[†],*

*Ignasi Fina[†]\*, and Josep Fontcuberta[†]\**

[†] Institut de Ciència de Materials de Barcelona (ICMAB-CSIC), Campus UAB, Bellaterra, Catalonia 08193, Spain

\*E-mail: ifina@icmab.es

\* E-mail: fontcuberta@icmab.cat

## Supporting Information 1

Figure S1a shows the amplitude PFM loops on the 2, 3 and 5 nm samples collected on top of the Pt electrode (where charging effects must be mitigated). Figure S1b shows the phase PFM loops showing 180° phase shift and butterfly loop near zeroing at the coercive field for the amplitude point. The top Pt electrodes (20 µm) have been mechanically lithographed using a diamond tip to define 2 x 2 µm² squares (Figure S1c).

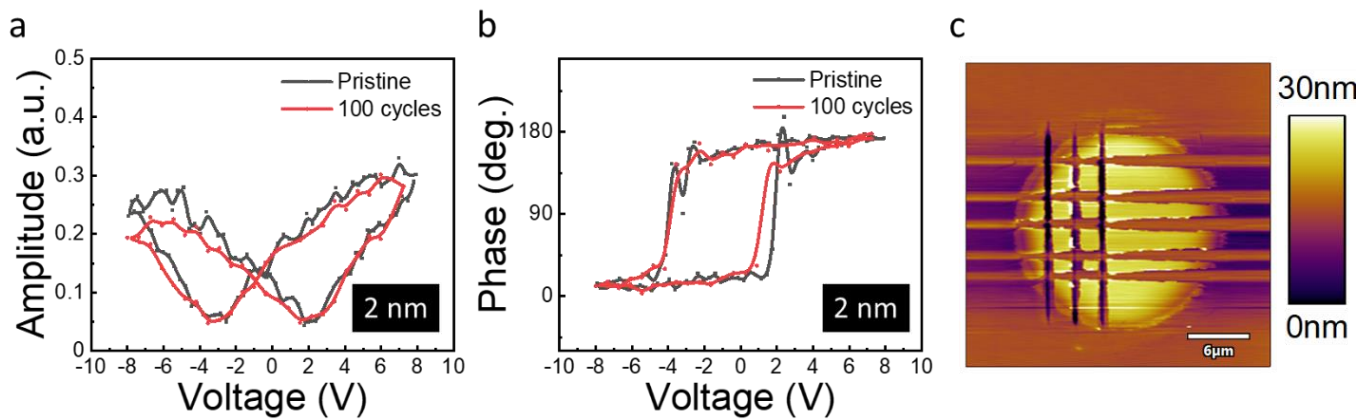

**Figure S1.** PFM (a) amplitude and (b) phase loops collected on pristine and $N_C$ = 100 cycled HZO (2 nm) junction on Pt electrode.

## Supporting Information 2

We show in Figure S2a,b the amplitude and phase loops recorded on 2 x 2 µm² electrodes, respectively, for the 3 and 5 nm samples. For comparison, the loops collected for the 2 nm samples are also shown. The top 2 x 2 µm² squares Pt electrodes have been mechanically defined using the diamond tip as shown in Figure S2c,d.

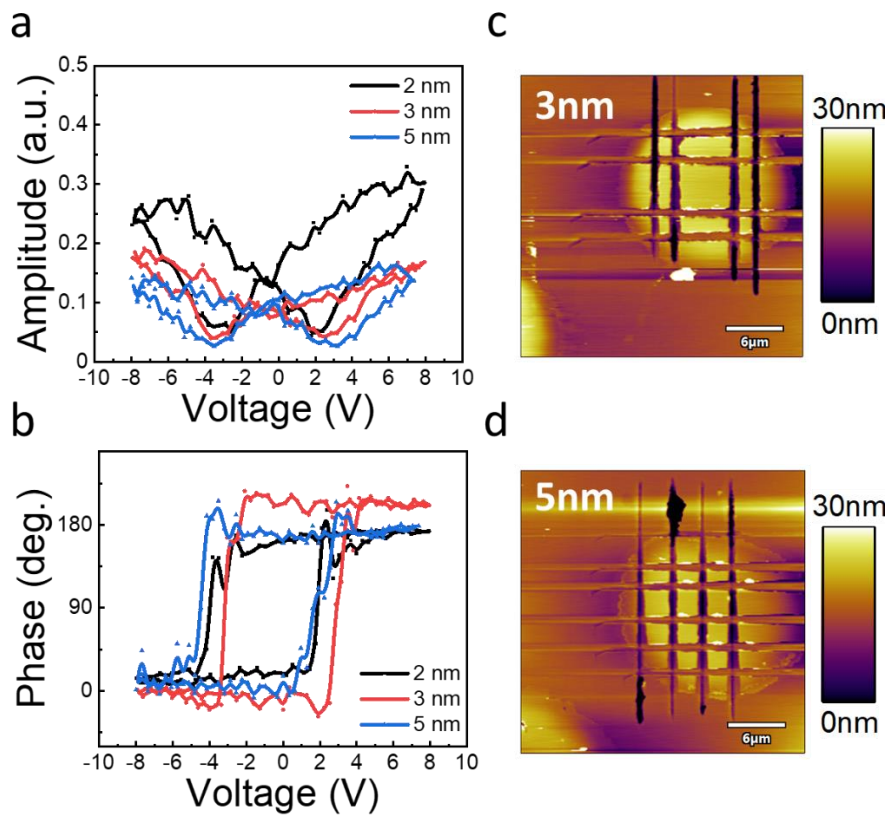

Figure S2. PFM (a) amplitude and (b) phase loops collected on a pristine area for the 2, 3 and 5 nm samples on 2 x 2 µm² Pt electrodes. (c,d) Topography of the scratched 2 x 2 µm² Pt electrodes for 3 and 5 nm samples, respectively.

## Supporting Information 3

Figures S3a,b shows the amplitude and phase PFM images, respectively, collected for the 2 nm sample in the pristine state. Figures S3c,d shows the amplitude and phase PFM images, respectively collected after applying positive and negative voltage three times to the whole probed surface. The same data is shown for the 3 nm sample in Figures S3e,f,g,h. The absence of important differences among all images indicates the absence of an important wake-up effect related to the emergence of ferroelectricity.

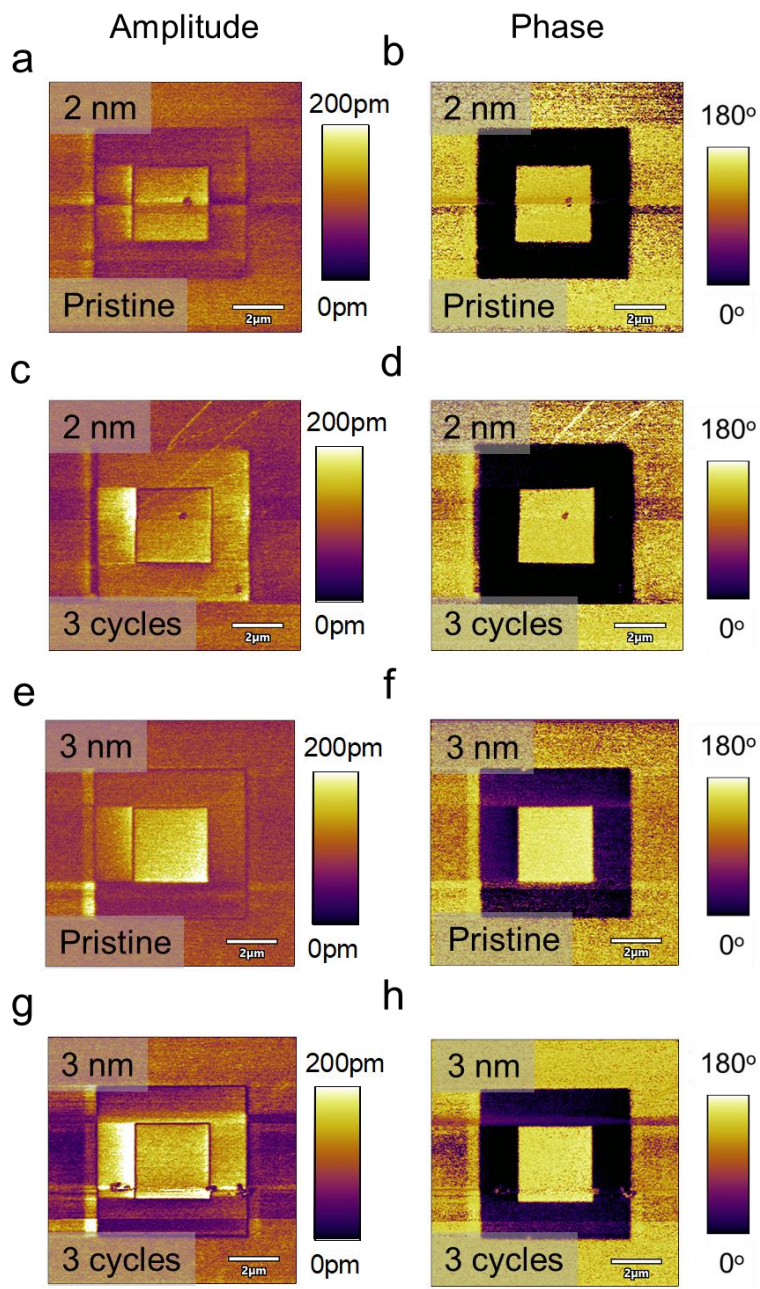

**Figure S3**. PFM (a) amplitude and (b) phase images collected on HZO (2 nm) sample in the pristine state. After the sample was cycled 3 times, the PFM c) amplitude and d) phase were collected again. PFM e) amplitude and f) phase images collected on HZO (3 nm) sample in the pristine state. After cycling 3 times, the g) amplitude and h) phase were measured.

## Supporting Information 4

Figure S4a shows XRD 2θ-χ frames and integrated scans of HZO/LSMO bilayers on GdScO$_3$ for samples with 2 nm sample described in the manuscript and plotted here for clarity. Equivalent data for the 3 nm and 5 nm films show no remarkable differences except for Pt (111) reflection due to the Pt electrodes shown in Figures S4b,c. Finer details of the evolution of HZO reflection with thickness can be better appreciated in Figure S4d, where we show the 2θ scan in a narrower angular range, where o-HZO (111) and m-HZO (002) reflection occur (vertical dashed lines). Data for all films display the obvious presence of o-HZO (111) reflection at 2θ = 30° and the absence of the m-HZO (002). For the thinnest film, the o-HZO (111) broadens due to size.

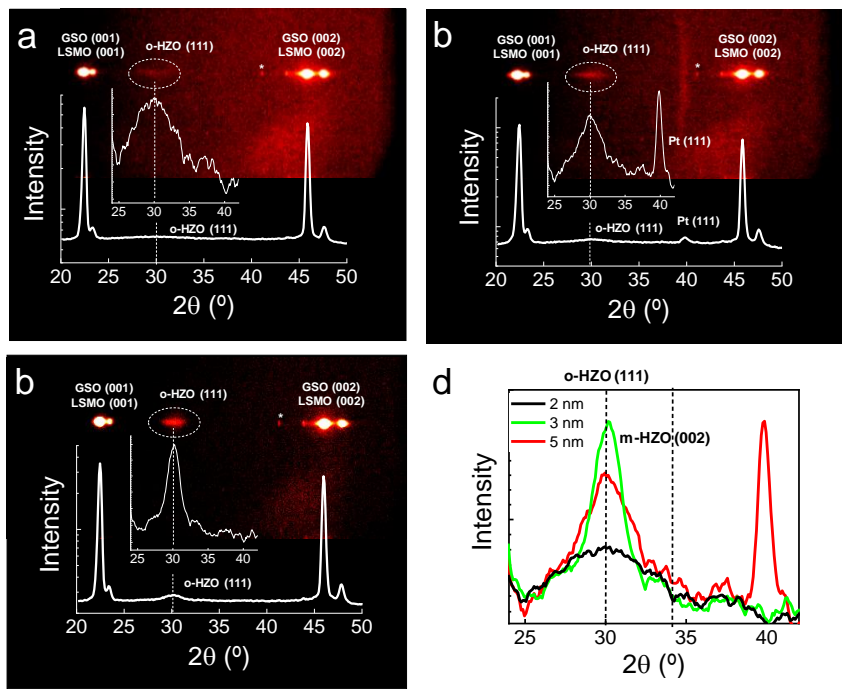

**Figure S4.** XRD 2θ-χ frames and integrated scans of HZO/LSMO bilayers on GdScO$_3$ for samples with a) 2, b) 3 and c) 5 nm HZO, respectively. d) Zoom around o-HZO (111) of the integrated scans along χ = ± 10° for the 2,3 and 5 nm sample.

## Supporting Information 5

Figure S5a and b are XRD 2θ scan and zoom, respectively, collected for HZO films with 2, 3 and 5 nm thickness. The most prominent peaks are ascribed to the LSMO bottom electrode and GSO substrate, as indicated. Several peaks marked with * correspond to non-filtered WK$_\beta$, CuK$_\beta$ and WK$_\alpha$ for increasing angles between 35 and 45°. The source light is not filtered to achieve the best signal-to-noise ratio. Similar conclusions to those extracted from data shown in Figure S4 hold here. Data for all films display the obvious presence of o-HZO (111) reflection and the absence of the m-HZO (002).

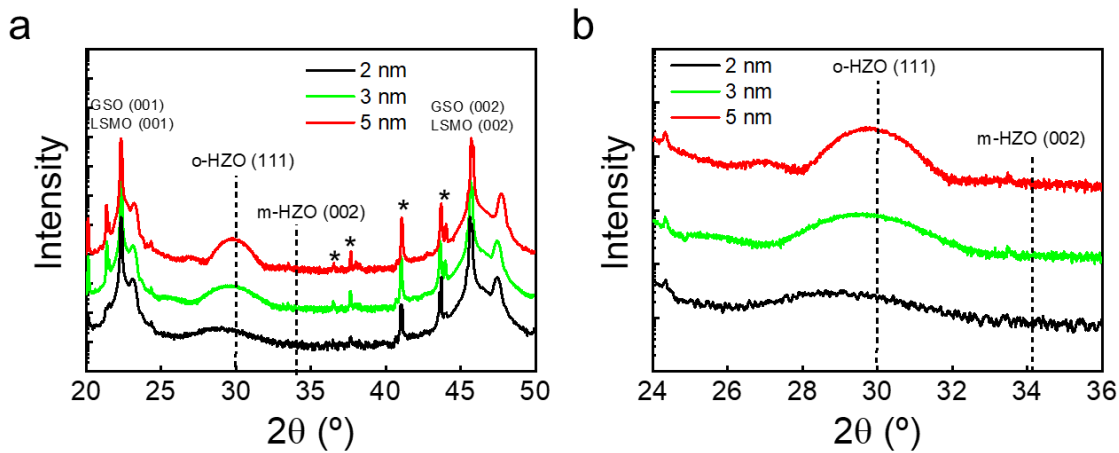

**Figure S5.** a) XRD 2θ scan in HZO films with 2, 3 and 5 nm thickness. b) Zoom around orthorhombic o-HZO (111) peak position.

## Supporting Information 6

In Figure S6a,d, the PUND raw data before subtraction and its zoom are shown. It can be observed that the leakage contribution is very large and originates, at the low voltage regime (up to 1 V), from the large tunneling current contribution as described in the main text. At the high voltage regime, other conduction mechanisms might also contribute.

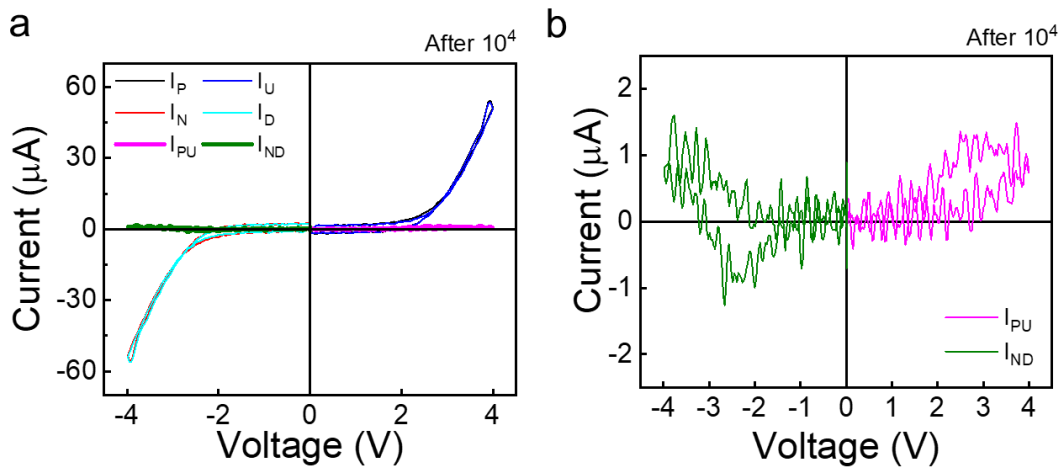

**Figure S6.** (a) Current versus voltage ($I_P$, $I_U$, $I_N$, $I_D$), containing leakage and polarization displacive currents, of the PUND measurement before PUND subtraction, measured at 5 kHz and up to 4 V for the 2 nm film. The thicker lines at very small current are the PUND results ($I_{PU}$, $I_{ND}$), containing the polarization reversal contribution after the corresponding subtraction. The PUND current is much smaller than the leakage, but switching is obvious, as shown in Figure 3b of the main text.

## Supporting Information 7

Data shown in Figure S7a,b does not allow observation of clear ferroelectric switching peaks for the pristine state. This might result from the much larger conductivity from the pristine state compared to the cycled one shown in Figure 3b that hinders the observation of the ferroelectric character of the sample. Note that these data are not conclusive regarding the ferroelectric nature of the junction.

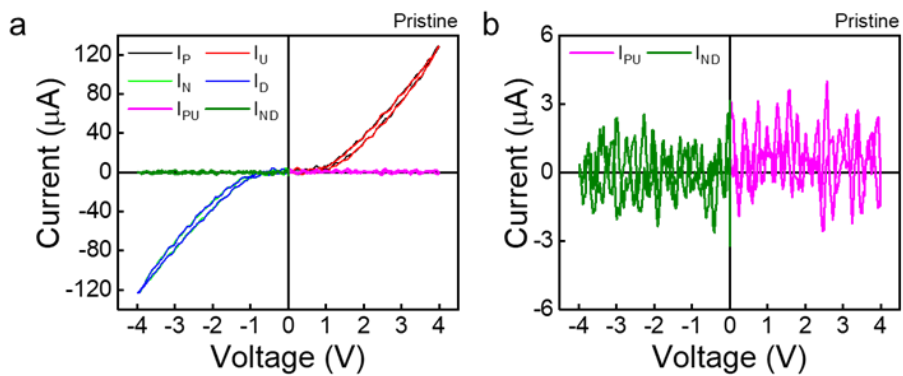

Figure S7. (a) Current versus voltage PUND loops before and after subtraction measured at 5 kHz and up to 4 V for the 2 nm film in the pristine state. (b) Zoom of Current versus voltage PUND loops after subtraction.

## Supporting Information 8

Data in Figure S8 shows $R(V_w)$ loops collected using 0.3 and 0.5 V reading voltages to that shown in Figure 3d using 1V. As expected, the resistance value is larger, the SNR smaller, but the ER is similar.

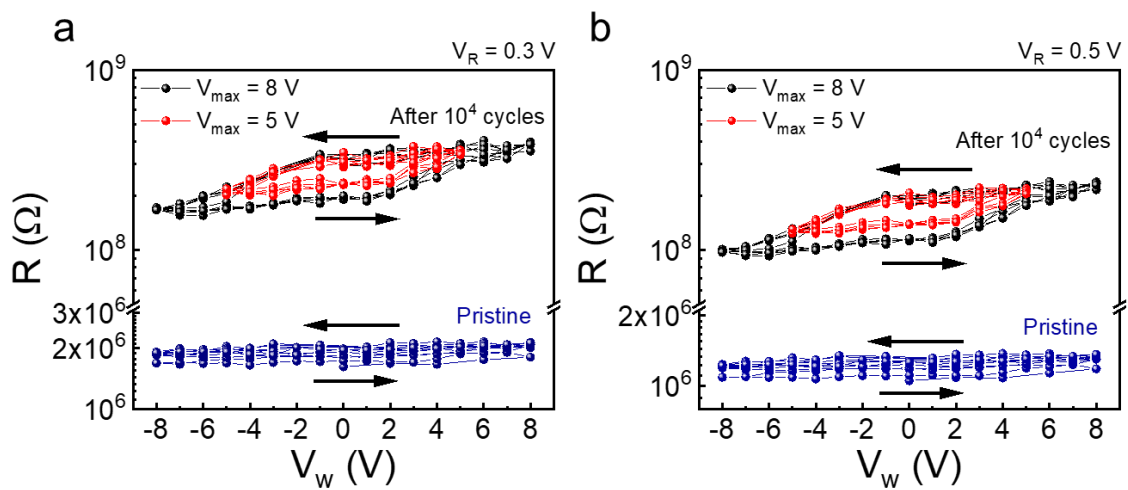

Figure S8. Dependence of junction resistance on the writing voltage measured in a junction after $10^4$ cycles using (a) 0.3 and (b) 0.5 V reading voltages.

## Supporting Information 9

From the comparison of the $dP/dV(V)$ loop (Figure S9a) and the $dR/dV_w(V)$ loop (Figure S9b), it can be observed that maxima take place at the same position indicating the connection between ER and ferroelectric switching.

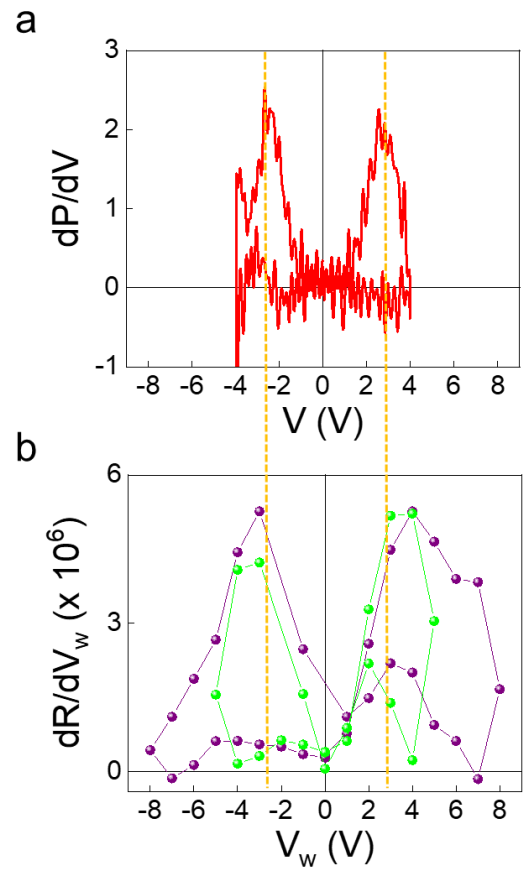

**Figure S9**. a) dP/dV obtaining by derivation of Figure 3b. b) dR/dV$_w$ obtained by derivation of Figure 3d. Peaks in both graphs coincide and indicate the coercive voltage.

# Supporting Information 10

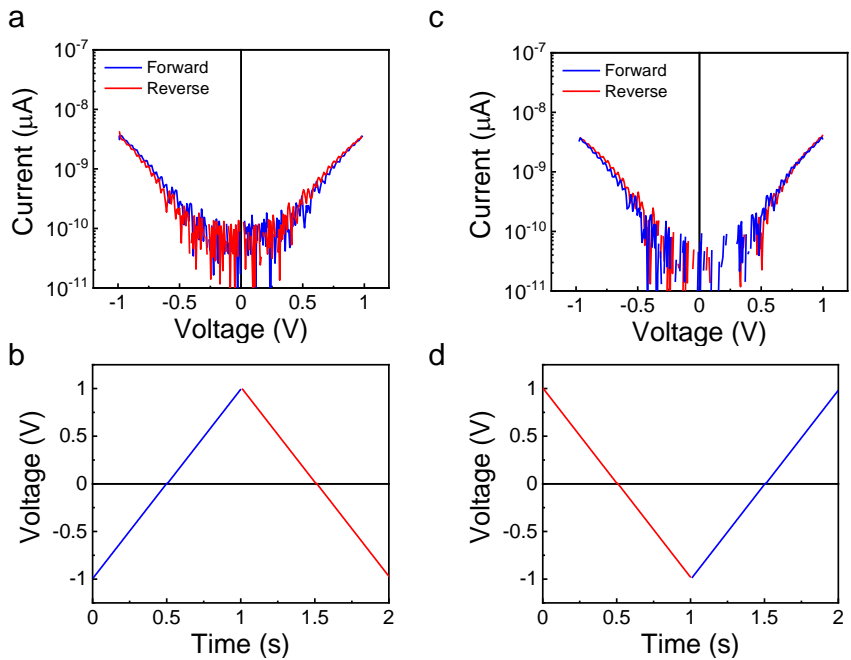

**Figure S10.** (a) Current-voltage hysteresis for (b) negative-positive-negative voltage polarity and (c,d) positive-negative-positive. The identical current-voltage curves discard resistive switching mediated by filament formation mechanism.

# Supporting Information 11

In order to evaluate the contribution of tunneling and leakage currents, the fittings of the I(V) curves from Figure 3b (main text) are shown here. The model used to fit the curves is the Brinkman model[91] associated with a parallel resistance, representing the leakage channel, as seen in Figure S11a. It is expected that as the leakage decreases its contribution with cycling, R increases its value.

One can compare the fitting taking into consideration only the Brinkman model or its association to R. Figure S11b shows the $\chi^2$ obtained from the fitting for both cases and the extracted values of R. Both models have a similar $\chi^2$ dependence with cycling and, as Figure S11c shows, neither of them can fit appropriately the region smaller than 1 V. In the case of the I(V) after $10^4$ cycles (Figure S11d), both models have similar fitting and $\chi^2$ value, which tunneling is now the major contribution for the current. However, for the pristine I(V), not even a Brinkman model with a variable parallel resistance is enough to describe the extremely leaky behavior in the junction. The extracted values of barrier heights, effective thickness (fixed), and parallel resistance are shown in Table S1.

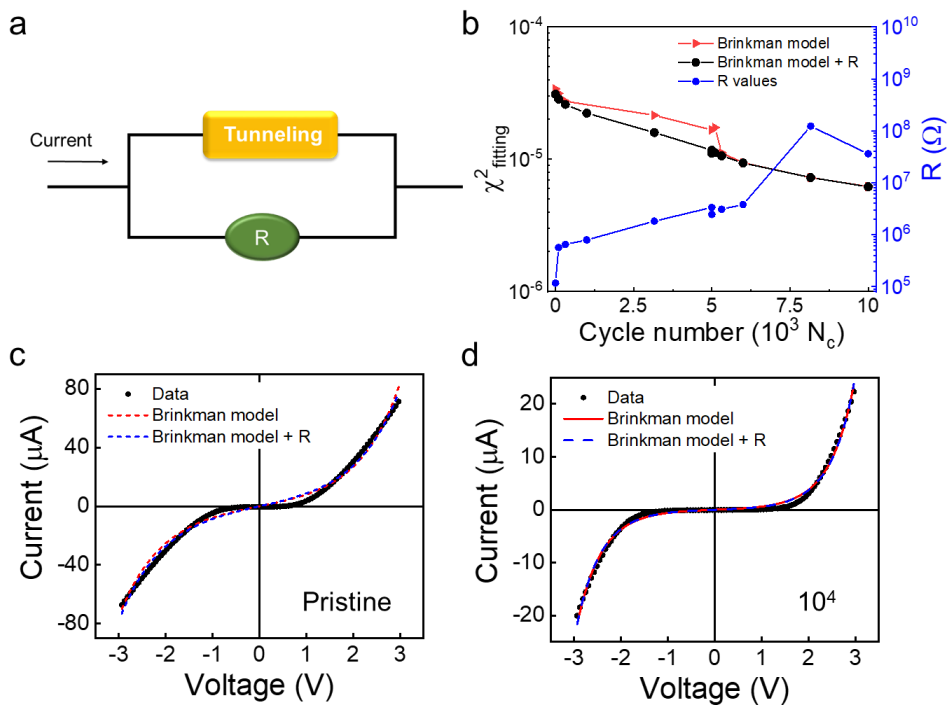

**Figure S11**. a) Equivalent circuit used for fitting. The current has two contributions: tunneling and parallel resistance. This resistance increases its value as the sample is cycled. b) $\chi^2$ extracted from fitting I(V) curves using only Brinkman model (black symbols) and using Brinkman model associated to a parallel resistance (red symbols). c) I(V) from a pristine device and fittings using both models. d) I(V) after $N_c = 10^4$ and fittings using both models.

| Cycle number | $t_{eff}$ (nm) | $\phi_{LSMO}$ (eV) | $\phi_{Pt}$ (eV) | R (Ω) |
|---|---|---|---|---|
| 1 | 3.15 | 3.17 | 3.19 | $3.6 \cdot 10^5$ |
| $5 \cdot 10^3$ | 3.15 | 3.46 | 3.40 | $3.3 \cdot 10^6$ |
| $10^4$ | 3.15 | 3.57 | 3.46 | $3.58 \cdot 10^7$ |

**Table S1.** Barrier heights ($\phi_{LSMO}$ and $\phi_{Pt}$), effective thickness ($t_{eff}$), and parallel resistance (R) value extracted from fitting.

## Supporting Information 12

The time constant of the circuit + sample was determined by applying a triangular voltage signal and measure the current response (Figure S12a) simultaneously. The voltage signal has 1 V amplitude and a frequency of 100 kHz. The current measured is delayed due to the time constant of the circuit, and its value can be calculated by fitting the current decay in time with an exponential curve $I = I_0 e^{-t/\tau_{RC}}$, where $I_0$ = amplitude and $\tau_{RC}$ = time constant.

In order to get a deeper insight on the time-dependent current I(t) response after a voltage pulse V(t), Figure S12b shows the current measured while applying rectangular V pulse trains: V > 0 pre-polarizing pulse and two consecutive negative pulses (V < 0) of 1 µs and 4 V as indicated.

The pre-polarizing pulse sets the polarization of HZO to a saturated state pointing down (towards LSMO). During this voltage step, displacive (charging the capacitor and driving of polarization) and non-displacive currents (mainly tunneling charges across the barrier) occur. The observation that the current is practically constant during the voltage plateau indicates that tunnel current dominates the response, overriding the small polarization-related, displacive current. The displacive current peak is apparent in the two subsequent negative pulses, followed by the tunneling-related current plateau. The fine details of I(t) can be better appreciated in Figure S12c, where the I(t) curves collected during the subsequent negative pulses are depicted, and its time-origin shifted to overlap. The mentioned displacive peaks appear at the voltage V(t) onset, followed roughly by the tunnel current plateau. More interesting, the amplitude of the displacive current is somewhat larger in the first V < 0 than at the second V < 0 pulse. This fine detail accounts for the polarization switching. Indeed, during the first negative pulse, the polarization is switched, and thus, polarization-switching displacive current adds to the capacitor charging current is measured: in contrast, only the latter contribution is present during the second negative pulse. Integration of the I(t) data leads to a polarization of ≈ 2 µC/cm$^2$, which is consistent with the data in Figure 3b obtained by PUND. More important here is that the polarization-switching current peak develops in a time scale shorter than 500 ns, which thus sets an upper limit for the polarization switching time. This value is comparable to the time constant from the (circuit + sample) system that limits the available time scale. We thus conclude that the intrinsic ferroelectric switching time faster than 500 ns. Reducing the time constant of the device by reducing its capacitance or the series resistance (due to the bottom electrode or circuit) should allow setting a lower limit for the switching time.

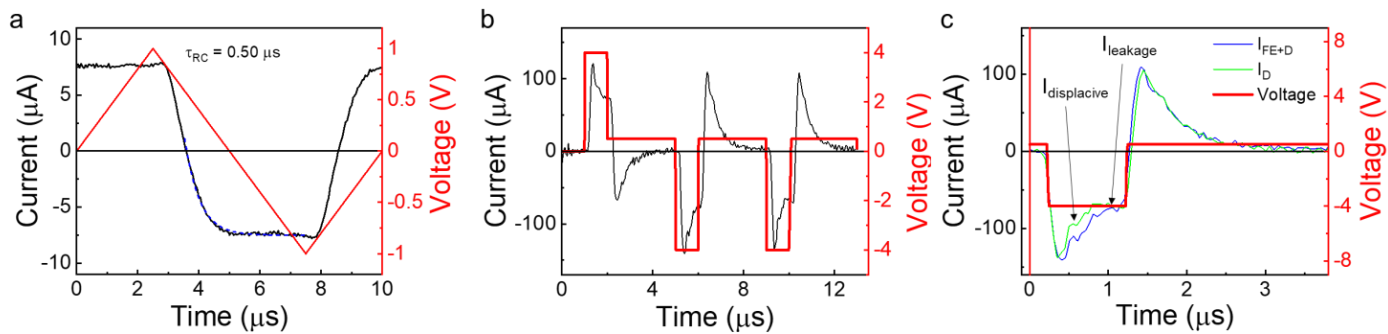

**Figure S12.** a) Current measured simultaneously with the application of triangular voltage pulse. The decay of current was fitted using an exponential equation, and the time constant ($\tau_{RC}$) was estimated as 0.50 µs. b) Complete train of voltage pulses applied and the measured current response. c) Current peaks and applied voltage for two consecutive negative pulses after pre-polarizing sample with $V_w$ = 4 V and duration = 1 µs.

## Supporting Information 13

In order to evaluate the resistance dependence on the number of pulses, the change in R with the number of pulses is fitted by $R_{p,d} = R_0^{p,d} + A_{p,d} e^{-(N_p - N_0^{p,d})/\tau_{p,d}}$, where p = potentiation, d = depression, $R_0$ = initial resistance, $N_p$ = number of pulses, $N_0$ = number of first positive (depression) or negative (potentiation) pulse and t = non-linearity of the effect.[70, 92-93] This dependence shows that the appropriate combination of voltage amplitude can achieve memristive behavior, duration and number of pulses (Figure S13). As close as the $\tau_p$ and $\tau_d$ are, smaller is the non-linearity of potentiation/depression effects, indicating a polar opposite behavior with the same number of pulses required. As a matter of fact, for application, this would be the ideal behavior. The results obtained for the fitting of data in Figure S13 are shown in Table S2. The asymmetry depending on voltage polarity is discussed in the main text.

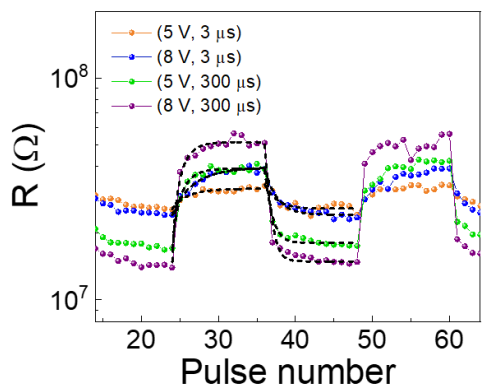

**Figure S13.** Potentiation/depression zoom showing results of the fitting.

| $V_w$ (V) | $\tau_w$ (µs) | $\tau_p$ | $\tau_d$ |
|---|---|---|---|
| 5 | 3 | 2.12 ± 0.48 | 1.47 ± 0.42 |
| 8 | 3 | 3.22 ± 0.51 | 1.45 ± 0.19 |
| 5 | 300 | 1.38 ± 0.15 | 0.81 ± 0.08 |
| 8 | 300 | 1.28 ± 0.22 | 0.66 ± 0.09 |

**Table S2.** Fitting parameters for each set ($V_w$, $\tau_w$) showed in Figure S13.

## Supporting Information 14

In Figure S14, the endurance of up to 130 cycles for pre-poling voltage of $V_w = \pm 4$ V and $\pm 8$ V is shown. It can be observed that the endurance for $V_w = \pm 4$ V is of the same order of magnitude as that reported previously in polycrystalline films.[71] For $V_w = \pm 8$ V, the recorded resistance state gradually increases, indicating some contribution of the discussed ionic motion contribution.

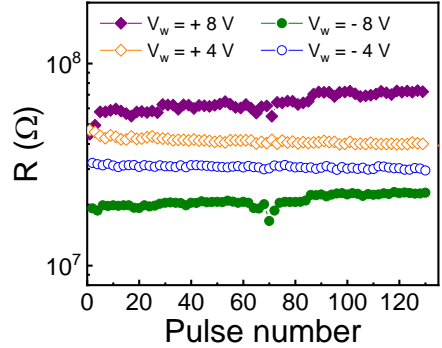

**Figure S14**. Resistance measured after indicated $V_w$ of opposite polarity applied continuously. The endurance for $V_w = \pm 4$ V and $V_w = \pm 8$ V were done in two different junctions at the 2 nm HZO sample after $N_c = 10^4$ cycles.

## Supporting Information 15

Figure S15a,b show PUND loops collected measured at 5 kHz for 3 and 5 nm films. Data in Figure S15c shows $P_r$ dependence on the thickness of the HZO films grown on LSMO/GSO from where it can be observed a typical peaky dependence. Figure S15d,e show PUND loops collected for increasing voltage measured at 5 kHz for 9 and 20 nm films

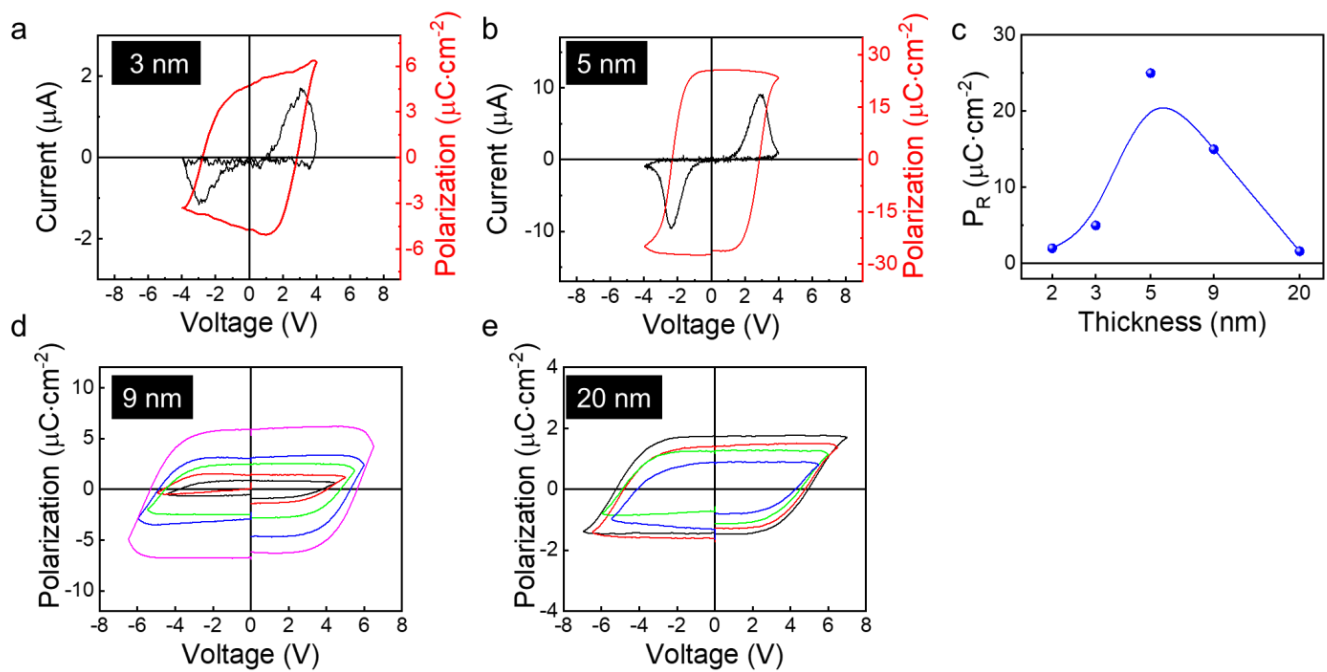

**Figure S15**. P(V) and I(V) measured with PUND in FTJ with a) 3 nm and b) 5 nm HZO. (c) $P_r$ dependence on thickness. P(V) loops collected with PUND for (a) 10 and (b) 20 nm HZO films.

## Supporting Information 16

In Figure S16a,b, we include resonance spectra for the 2, 3 and 5 nm samples collected, placing the tip on the surface and on the 2 x 2 µm² electrodes, respectively. It can be observed that the peak amplitude is proportional to the sample thickness. As the resonance amplitude is proportional to the sample's thickness and the $d_{33}$, we have normalized the peak amplitude to the sample thickness to remove its contribution and show in Figure S16c. As the excitation voltage is 200 mV in all the experiments, the normalized amplitude is comparable. In Figure S16c, it can be observed that it increases with thickness. Thus, it can be concluded that piezoresponse and polarization increase with thickness in the explored thickness range.

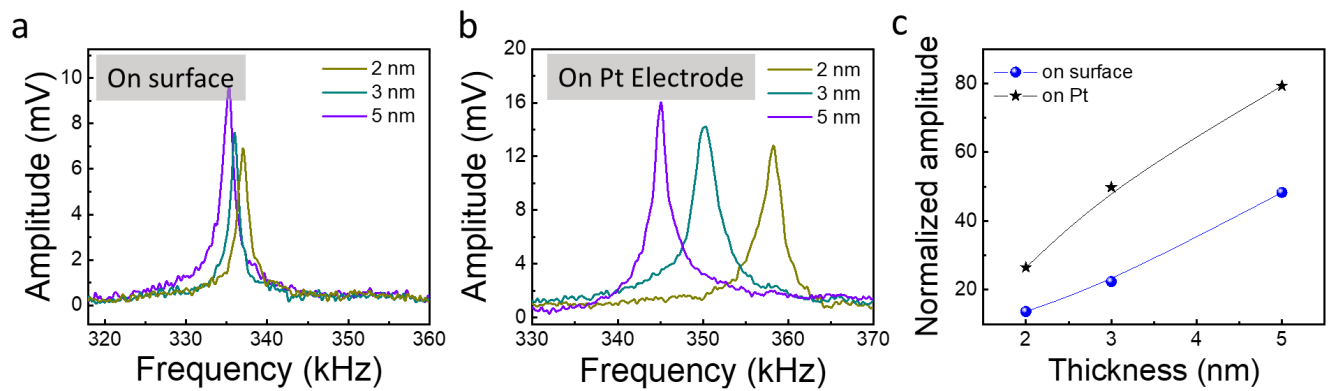

Figure S16. Resonance spectra collected for 2, 3 and 5 nm samples placing the tip (a) on the surface and (b) on 2 x 2 µm² Pt electrodes. (c) Dependence of the normalized to the thickness amplitude of the resonance peak as a function of sample thickness.

## Supporting Information 17

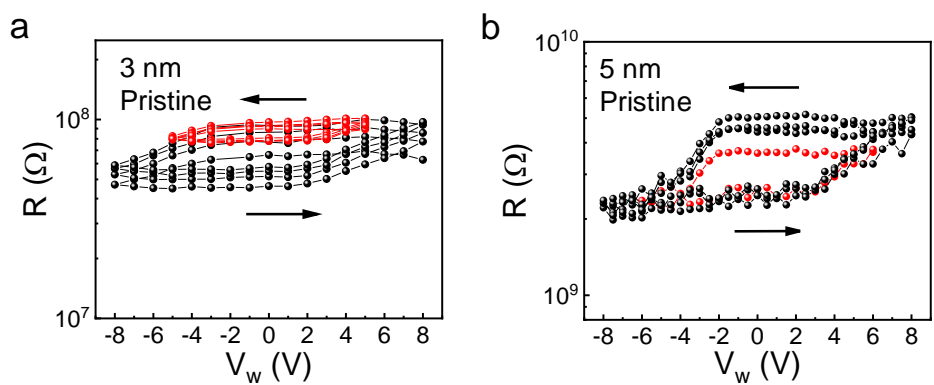

**Figure S17.** Resistance dependence on the external voltage pulse amplitude ($V_w$) measured for pulses with $\tau_w$ = 300 µs obtained before cycling in HZO junction with a) 3 nm and b) 5 nm thickness.